\documentclass[letterpaper,conference]{IEEEtran}

\usepackage{graphicx}
\usepackage[utf8]{inputenc} 
\usepackage[T1]{fontenc}
\usepackage{url}
\usepackage{ifthen}
\usepackage{cite}
\usepackage{color}
\usepackage{comment}

\usepackage{cite}

\ifCLASSINFOpdf
\else
\fi
\usepackage{array}
\usepackage{graphicx}
\usepackage{amsmath}
\usepackage{amssymb}
\usepackage{boldline}
\usepackage{bm}
\usepackage{bbm}
\usepackage{algorithmicx}
\usepackage{algorithm}
\usepackage{algpseudocode}
\usepackage{array}
\newcolumntype{P}[1]{>{\centering\arraybackslash}p{#1}}
\newcolumntype{R}[1]{>{\arraybackslash} \hfill p{#1}}
\usepackage{booktabs}
\usepackage{multirow}
\usepackage{makecell}
\usepackage{breqn}

\usepackage{tikz}
\usetikzlibrary{shapes.geometric, arrows}

\begin{document}
%
\title{ELF Codes: Concatenated Codes with an Expurgating Linear Function as the Outer Code }
%
\author{Richard Wesel\IEEEauthorrefmark{1},
Amaael Antonini\IEEEauthorrefmark{1},
Linfang Wang\IEEEauthorrefmark{1},
Wenhui Sui\IEEEauthorrefmark{1},
Brendan Towell\IEEEauthorrefmark{1},
and
Holden Grissett\IEEEauthorrefmark{1}\\

\IEEEauthorblockA{\IEEEauthorrefmark{1}Department of Electrical and Computer Engineering, University of California, Los Angeles, Los Angeles, CA 90095, USA}

Email: \{wesel, amaael, lfwang, wenhui.sui, brendan.towell, holdengs\}@ucla.edu 
}


\maketitle

\begin{abstract}
An expurgating linear function (ELF) is an  outer code that disallows low-weight codewords of the inner code.  
ELFs can be designed either to maximize the minimum distance or to minimize the codeword error rate (CER) of the expurgated code. A list-decoding sieve can efficiently identify ELFs that maximize the minimum distance of the expurgated code.  For convolutional inner codes, this paper provides analytical distance spectrum union (DSU) bounds on the CER of the concatenated code. 

For short codeword lengths, ELFs transform a good inner code into a great concatenated code.    For a constant message size of $K=64$ bits or constant codeword blocklength of $N=152$ bits, an ELF can reduce the gap at CER $10^{-6}$ between the DSU and the random-coding union (RCU) bounds from over 1 dB for the inner code alone to 0.23 dB for the concatenated code.  The DSU bounds can also characterize puncturing that mitigates the rate overhead of the ELF while maintaining the DSU-to-RCU gap. List Viterbi decoding guided by the ELF achieves maximum likelihood (ML) decoding of the concatenated code with a sufficiently large list size. The rate-$K/(K+m)$ ELF outer code reduces rate and list decoding increases decoder complexity.  As SNR increases, the average list size converges to 1 and  average complexity is similar to Viterbi decoding on the trellis of the inner code. For rare large-magnitude noise events, which occur less often than the FER of the inner code, a deep search in the list finds the ML codeword.

\end{abstract}



{\let\thefootnote\relax\footnote{{This research is supported by National Science Foundation (NSF) grant CCF-2008918. Any opinions, findings, and conclusions or recommendations expressed in this material are those of the authors and do not necessarily reflect views of NSF.}}}

%
\IEEEpeerreviewmaketitle

\section{Introduction}
\label{sec:Intro}
Expurgating a code decreases the number of message symbols without changing the length \cite{Blahut2003}.  Expurgation strengthens a code by removing weaker codewords at the expense of a slight reduction in rate.  For example, in his proof of channel capacity, Gallager \cite{Gallager1965} removes the half of the randomly selected codewords for which error probability is largest to bound maximum rather than average probability of error.  For a linear code, where the minimum distance $d_{\text{min}}$ is the minimum weight of a nonzero codeword \cite{Blahut2003}, expurgation that removes the lowest-weight codewords will increase $d_{\text{min}}$  and thus reduce the probability of a codeword error. 

Practical expurgation requires a function, i.e. an outer code, that selects which codewords to remove. This paper develops the paradigm of using an expurgating linear function (ELF) to remove the lowest weight codewords of a linear code or, more generally, to remove codewords so as to minimize a union bound on codeword error rate (CER). The improved performance comes at the expense of a reduction in rate by the rate-$K/(K+m)$ ELF outer code, but well-designed ELFs move CER performance closer to an appropriately rate-adjusted random coding union (RCU) bound \cite{Polyanskiy,Font-Segura2018}.

Lou {\em et al.} in \cite{Lou2015} significantly improved error detection performance over traditional cyclic redundancy codes (CRCs) used with zero-terminated convolutional codes by specifically designing CRCs that remove low-weight codewords and thus reduce the undetected error rate.  Subsequent papers \cite{YangGlobecom2018,YangGlobecom2019,LiangGlobecom2019,Yang2020,YangTCOM2022,KingICC2022,KingGlobecom2022,SongGlobecom2022,KingThesis,SuiICC2022,WangICC2022,Wang2023} used this approach to improve the minimum distance or CER performance of the overall concatenated code, which is decoded using list decoding. These subsequent papers characterized the new designs  as CRCs, but we call them ELFs since they are not constrained to be a cyclic code of any particular length and their primary function is expurgation rather than error detection.

For a convolutional inner code with an ELF as the outer code,  ML decoding is achieved by serial or parallel list Viterbi decoding with a sufficiently large list size \cite{Seshadri1994}.  As observed in \cite{YangTCOM2022}, as SNR increases, the expected list size converges to one.  The average list size is often small at the desired CER.  

BCH codes \cite{YangGlobecom2019} and legacy codes  that include CRCs can be reconsidered as ELF codes and decoded using list decoding to decrease CER. Schiavone {\em et al.} \cite{Schiavone2022} provide a compelling example. The ELF paradigm  applies to any inner code.  Building on \cite{TalIT2015}, ELFs designed to maximize the minimum distance of the expurgated polar code are described in \cite{KingICC2022,KingArXiv2023}.  An ELF for a trellis code appears in \cite{Wang2023}. This paper focuses on ELFs concatenated with tail-biting convolutional codes (TBCCs) \cite{Ma1986} as an example.

\subsection{Contributions}

This paper presents generating function techniques for distance spectrum union (DSU) upper bounds on the CER of TBCCs concatenated with ELFs under maximum-likelihood (ML) decoding. These bounds are extended to include puncturing for rate compatibility. Such bounds can characterize the CER of every ELF for a given redundancy $m$ and can be used to select the ELF (and the puncturing) that minimizes CER.  Alternatively, this paper provides a sieve approach that uses list decoding from the zero-message codeword to identify the ELF that maximizes the minimum distance of the expurgated code.  The DSU bounds show that an ELF can improve the gap between the DSU and RCU bounds from over 1 dB for the inner code alone to 0.23 dB for the concatenated code.

\subsection{Organization}

Sec. \ref{sec:AnalyticalBounds} presents (DSU) upper bounds on the CER of convolutional inner codes concatenated with ELF outer codes with or without puncturing. Sec. \ref{sec:ListDecodingSieve} presents a  sieve approach to identify the ELF that maximizes the minimum distance of the expurgated code.  As an example, the best ELFs for expurgating a  (152, 76) block code created from a $\nu=8$  tail-biting convolutional code are identified for ELF redundancies of $0\le m\le 12$. Sec. \ref{sec:Example} uses the tools of Sec. \ref{sec:AnalyticalBounds} and Sec. \ref{sec:ListDecodingSieve} to design an example punctured concatenation of an ELF and a tail-biting convolutional code.  Sec. \ref{sec:Conclusions} concludes the paper.

\section{Distance Spectrum Union Bounds}
\label{sec:AnalyticalBounds}

Generating functions provide upper bounds on the bit error rate \cite{ViterbiTCOM1971} and the CER \cite{Lou2015,SongGlobecom2022} of convolutional codes.  This section reviews the generating function approach to computing distance spectrum union (DSU) bounds on the CER of block codes constructed using convolutional codes and describes how such bounds may be applied to zero-terminated  and tail-biting convolutional codes with ELFs and with puncturing.

\subsection{DSU Bounds for Zero Termination and Tail Biting}
Consider an $(N,K)$ binary block code transmitted over the binary input AWGN channel where $+\sqrt{E_s}$ is transmitted for binary 0 and $-\sqrt{E_s}$  is transmitted for binary 1.  Let $A_w$ be the number of codewords with Hamming weight $w$.  A DSU bound on codeword error rate $P_{cw}$ is shown below:
\begin{equation}
  P_{cw} \le \sum_{w=d_{\text{min}}}^N A_w Q \left (\sqrt{\frac{wE_s}{\sigma^2}} \right )  . \label{eq:unionbound}
\end{equation}

Define the weight enumerator polynomial as
\begin{equation}
 A(w) = \sum_{w=d_{\text{min}}}^N A_w W^w .
\end{equation}
Using the bound $Q(\sqrt{x+y}) \le Q(\sqrt{x}e^{-y/2})$ \cite{ViterbiTCOM1971}, a slightly looser bound than \eqref{eq:unionbound} may be computed from the weight enumerator polynomial as follows:
\begin{equation}
  P_{cw} \le  Q \left (\sqrt{\frac{d_{\text{min}}E_s}{\sigma^2}} \right ) e^{\frac{d_{\text{min}}E_s}{2\sigma^2}} A \left ( e^{-\frac{E_s}{2\sigma^2}}\right )   . \label{eq:WEPunionbound}
\end{equation}

A transition matrix $T(W)$ facilitates computation of $A(W)$ for either a zero-terminated or tail-biting  convolutional code.  For the example of a rate $1/n$ convolutional code with $\nu$ memory elements, $T(W)$ is an $s \times s$ matrix, where  $s = 2^{\nu}.$

The entry of $T(W)$ at row $j$ and column $i$ represents the transition from state $i$ to state $j$ in the convolutional encoder. If there is no transition from state $i$  to state $j$, then the entry is zero. Otherwise, the entry is $W^{w_{i,j}}$ where $w_{i,j}$ is the Hamming weight of the $n$-bit symbol transmitted for that state transition.

Define the $i^{\text{th}}$ basic row vector $e_i$ as a length-$s$ vector of all-zeros except a one in position $i$.  We will index the entries in both $T(W)$ and $e_i$ from zero to $s-1$. For example, with $\nu = 2$ and thus $s=4$, $e_1 = \begin{bmatrix}
    0&1&0&0
\end{bmatrix}$. 

For an $(N=n(K+\nu),K)$ block code implemented by sending $K$ information bits through a $\nu$-state, rate-$1/n$, zero-terminated convolutional code, 
\begin{equation}
    A(W) = e_0 T(W)^{K+\nu}e_0^T -1 . \label{eq:DSU-ZTCC}
\end{equation}
Similarly, for an $(N=nK,K)$ block code implemented by sending $K$ information bits through a $2^\nu$-state tail-biting convolutional code, 
\begin{equation}
    A(W) = \sum_{i=0}^{2^{\nu} -1}e_i T(W)^{K}e_i^T -1 .
\end{equation}

\subsection{DSU Bound for a Convolutional Code with an ELF}
A degree-$m$ ELF $E(x)$ can be concatenated with a zero-terminated rate-$1/n$ convolutional code with encoder polynomial matrix $G(x)$.  The $m$ ELF redundancy bits reduce the rate from $\frac{K}{(K+\nu)n}$ to $\frac{K}{(K+\nu+m)n}$. Eq. \eqref{eq:DSU-ZTCC} can be applied to the $T(W)$ with $s=2^{(\nu + m)}$ for the zero-terminated convolutional code $E(x)G(x)$ so that 
\begin{equation}
    A(W) = e_0 T(W)^{K+\nu+m}e_0^T -1 . \label{eq:DSU-ELF-ZTCC}
\end{equation}

\begin{figure}[t]
\centering
    \includegraphics[width=21pc]{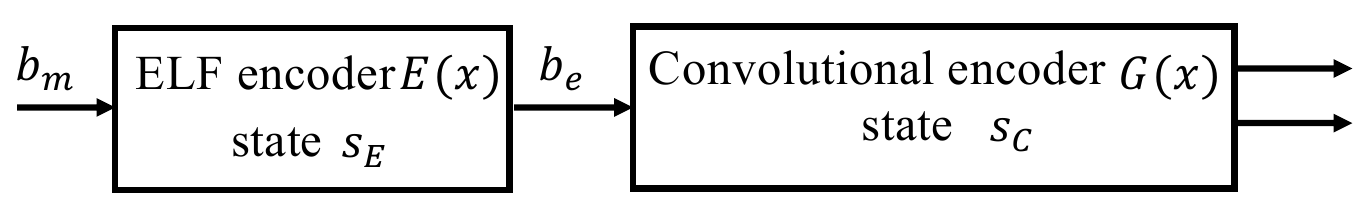}
\caption{Convolutional encoder $G(x)$ with an ELF $E(x)$ as an outer code.}
    \label{fig:ELF-CC}
\end{figure}

For the case of an ELF $E(x)$ concatenated with a  rate-$1/n$ tail-biting convolutional code, the rate reduction is from $\frac{K}{Kn}$ to $\frac{K}{(K+m)n}$. To derive the DSU bound, separately consider the state $s_E$ of the ELF encoder in Fig. \ref{fig:ELF-CC} and the state $s_C$ of the tail-biting convolutional encoder in Fig. \ref{fig:ELF-CC}.   Note that
\begin{align}
0 \le &s_E \le 2^m-1\\
0 \le &s_C \le 2^{\nu} -1
\end{align}

Define the overall encoder state of the concatenated code as   $s = s_C + 2^{\nu} \times s_E$.  To compute an entry in $T(W)$, the following computations are performed (referencing Fig. \ref{fig:ELF-CC}): \\ 1) The message input bit $b_m$ is processed by the ELF encoder with the origin ELF state $s_E^{(o)}$ to compute the destination ELF state $s_E^{(d)}$ and ELF output bit $b_e$. 2) The ELF output bit $b_e$ is the input to the convolutional encoder which updates its origin state $s_C^{(o)}$ to the destination state $s_C^{(d)}$ and produces an $n$-bit output with Hamming weight $w_{i,j}$.  3) The weight term $W^{w_{i,j}}$ is written to $T(W)$ at row $j$ and column $i$ where $j = s_C^{(d)} + 2^{\nu} \times s_E^{(d)}$ and $i= s_C^{(o)} + 2^{\nu} \times s_E^{(o)}$. Using this $T(W)$, the weight enumerator polynomial for an $(N=2\times(K+m),K)$ block code implemented by sending $K$ information bits through an outer ELF code with $2^m$ states and encoding the ELF output with a $2^\nu$-state tail-biting convolutional code is 
\begin{equation}
    A(W) = \sum_{i=0}^{2^{\nu} -1}e_i T(W)^{K+m}e_i^T -1 . \label{eq:DSU-ELF-TBCC}
\end{equation}
In \eqref{eq:DSU-ELF-TBCC}, $T(W)$ is an $2^{m+\nu} \times 2^{m+\nu}$ matrix, but the only paths that contribute to $A(W)$ are the $2^{\nu}$ paths that start and end at the same value of $s_C$ and that start and end with $s_E=0$.

\begin{figure}[t]
    \centering
\includegraphics[width=21pc]{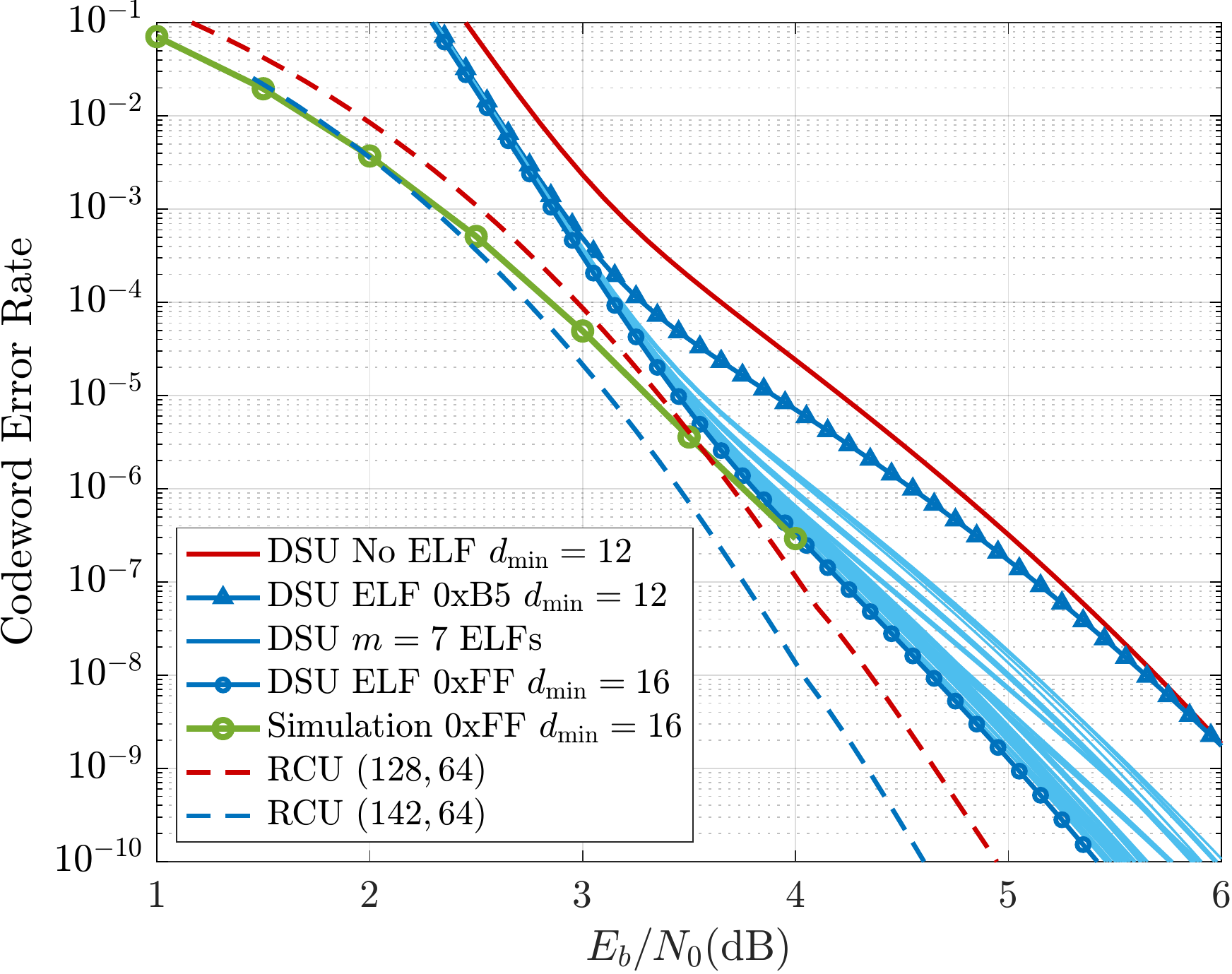}
    \caption{DSU bounds for the $\nu=8$ tail-biting convolutional code with $K=64$ message bits with no ELF (red) and with each possible $m=7$ ELF (blue). Also shown is a simulation of list Viterbi decoding of the best ELF 0xFF (green) and, for reference, the (142,64) and (128,64) RCU bounds (dashed).}
    \label{fig:allCRCv8m7K64}
\end{figure}

Fig. \ref{fig:allCRCv8m7K64} shows the DSU bounds computed according to \eqref{eq:DSU-ELF-TBCC} and \eqref{eq:WEPunionbound} for all of the 64 possible (142, 64) codes resulting from an $m=7$ ELF concatenated with the $\nu=8$ tail-biting convolutional code (561,753), where 561 is octal for  $1+x^4+x^5+x^6+x^8$. The RCU bound for (142,64) codes is shown for comparison.  For reference, the DSU bound for the (128,64) code that results from using the tail-biting convolutional code without an ELF and its corresponding RCU bound are also shown.  The CER curve from simulating list decoding of the tail-biting code used with the best ELF 0xFF shows that this DSU bound is tight for CER $\le 10^{-6}$.

Fig. \ref{fig:allCRCv8m7K64} shows how the DSU bound on CER varies with the choice of ELF.  The best ELF is 0xFF whose DSU bound is 0.35 dB from the (142,64) RCU bound at CER=$10^{-6}$.  The worst ELF is 0xB5 which is {1.10 dB} from the RCU bound at CER=$10^{-6}$.  The original (128,64) TBCC is 1.05 dB from the (128,64) RCU bound.  Thus, in this case a well-designed 7-bit ELF reduced the gap from the RCU bound by 0.65 dB while the poorest choice had a slightly larger gap.

\begin{figure}[t]
    \centering
\includegraphics[width=21pc]{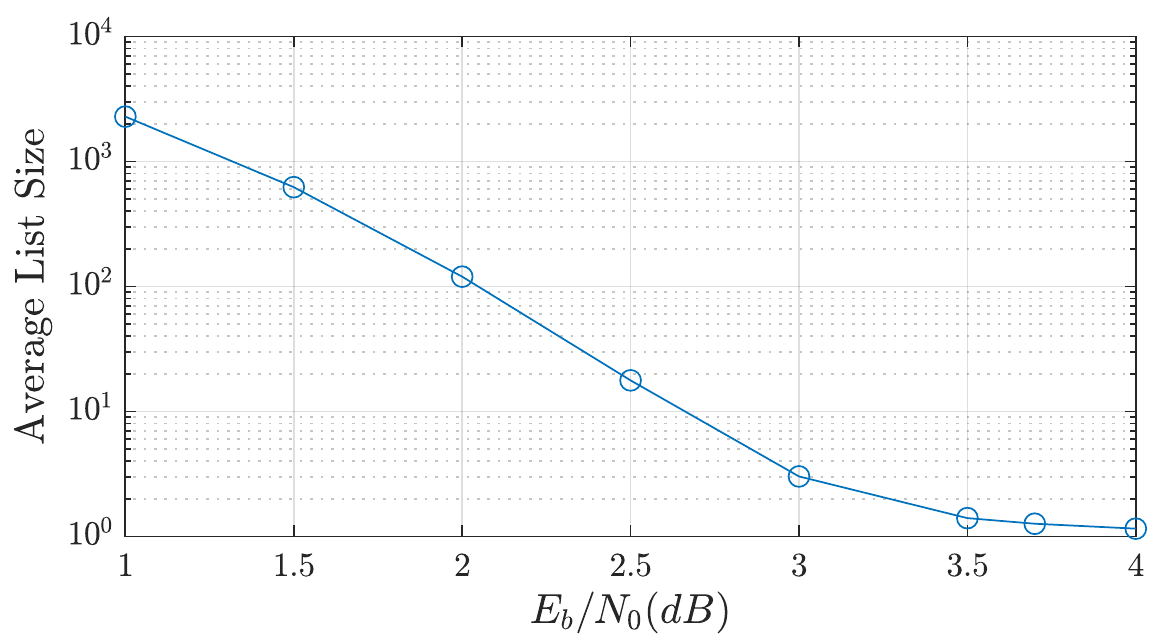}
    \caption{Average list size vs. $E_b/N_0$ for the list Viterbi decoding simulation of a $\nu=8$ TBCC concatenated with ELF 0xFF shown in Fig.\ref{fig:allCRCv8m7K64} }.
    \label{fig:mean_list_sizes}
\end{figure}

Fig \ref{fig:mean_list_sizes} shows average list size as a function of $E_b/N_0$ for the list decoding simulation shown in Fig. \ref{fig:allCRCv8m7K64}. The maximum list size was set at $2^{20}$, but the average list size is 1.26 at $E_b/N_0$=3.7 dB, where the CER is  $1.1 \times 10^{-6}$.  Thus the average complexity is similar to regular Viterbi decoding on a 256-state trellis but the gap to the RCU bound is only 0.35 dB.

\subsection{DSU Bound for Punctured Convolutional Code with ELF}
\label{sec:puncturingDSU}

This section extends the DSU bounding technique to handle puncturing.  The techniques below can be applied to any puncturing scheme. For simplicity of exposition, our description only considers puncturing at most one of the $n$ bits in each convolutional encoder symbol.  Let $p$ be the puncturing index with range $0\le p \le n$, which indicates either that no bit is punctured ($p=0$) or which bit is punctured $1\le p \le n$.

Let $T_p(W)$ be the transition matrix of a trellis stage with puncturing according to puncturing index $p$.  The matrix $T_0(W)$ is the same as the $T(W)$ discussed above.  The matrices $T_p(W)$ for $1\le p \le n$ have the power of $W$ reduced by one for the entries where the  $p^{\text{th}}$ output bit, i.e. the punctured bit, is a 1. With this description of punctured transition matrices $T_p(W)$, let $p_i$ be the puncturing index that describes the puncturing of the $i^{\text{th}}$ trellis stage.  The weight enumerator polynomial for the  block code implemented by sending $K$ information bits through an outer ELF code with $2^m$ states and then a $2^\nu$-state tail-biting convolutional code and then puncturing some number of bits is expressed as follows:
\begin{equation}
    A(W) = \sum_{i=0}^{2^{\nu} -1}e_i \prod _{i=1}^{K+m}T_{p_i}(W)e_i^T -1 . \label{eq:DSU-PUNCT-ELF-TBCC}
\end{equation}

Often, puncturing is designed according to a periodic pattern.  If the period includes $q$ trellis stages, then we can define a transition matrix $T_{\pi}(W)$ for the puncturing period as follows:
\begin{equation}
    T_{\pi}(W)= \prod _{i=1}^{q}T_{p_i}(W) \, .
\end{equation}
If the $K+m$ trellis stages are an integer number of puncturing periods, the weight enumerator of the punctured tail-biting convolutional code with an ELF can be expressed as
\begin{equation}
    A(W) = \sum_{i=0}^{2^{\nu} -1}e_i T_{\pi}(W)^{(K+m)/q}e_i^T -1 . \label{eq:DSU-PPUNCT-ELF-TBCC}
\end{equation}
\section{A List Decoding Sieve to find the best ELF}
\label{sec:ListDecodingSieve}
\begin{table}[t] 
\centering
\caption{Best ELFs $E(x)$ for $m=0$ to $m=12$ redundancy bits\\ that maximize minimum distance for the $\nu=8$ TBCC (561,753)\\ for $K=64$ message bits, $N=2\times(64+m)$ transmitted bits and for $N=2\times76$ transmitted bits, $K=76-m$ message bits.}
\label{tab:ELFsK64andN76}
\begin{tabular}{P{0.3cm} P{1.4cm} P{0.6cm} P{0.6cm} | P{1.4cm} P{0.6cm} P{0.6cm} }
\hline
\clineB{1-7}{1.2}\\ [\dimexpr-\normalbaselineskip+2pt]
& \multicolumn{3}{c|}{$K=64, N=2\times(64+m)$}& \multicolumn{3}{c}{$N=2\times76, K=76-m$}\\ 
$m$  &$E(x)$& $d_{\text{min}}$ & $A_{d_{\text{min}}}$ &$E(x)$ & $d_{\text{min}}$  & $A_{d_{\text{min}}}$ \\
 \hline 
$0$ & 0x1 & 12 & 704 & 0x1 & 12 & 836\\
$1$ & 0x3 & 12 & 260 & 0x3 & 12 & 304 \\
$2$ & 0x5 & 12 & 66 & 0x5 & 12 & 76\\
$3$ & 0xF & 12 & 4 & 0xF & 14 & 380\\
$4$ & 0x11 & 14 & 68 & 0x11 & 14 & 76\\
$5$ & 0x33 & 14 & 11 & 0x33 & 14 & 4\\
$6$ & 0x7F & 16 & 210 & 0x55 & 14 & 2\\
$7$ & 0xFF & 16 & 86 & 0x81 & 16 & 24\\
$8$ & 0x1AB & 18 & 360 & 0x195 & 16 & 6\\
$9$ & 0x301 & 18 & 146 & 0x325 & 18 & 297\\
$10$ & 0x4F5 & 18 & 17 & 0x53D & 18 & 21\\
$11$ & 0x9AF & 20 & 300 & 0xE0D & 18 & 2\\
$12$ & 0x1565 & 20 &47& 0x1565 & 20 &47\\
 \hline
\clineB{1-7}{1.2}
\end{tabular}
\end{table}

\begin{table}[t] 
\centering
\caption{Expurgated distance spectra for $m=0$ (no expurgation) to $m=12$ for the $\nu=8$ (N=152,K=76) tail-biting convolutional mother code described by polynomial (561,753).}
\label{tab:ExpurgatedDS}
\begin{tabular}{P{0.2cm} P{1.0cm}| P{0.6cm} P{0.6cm}  P{0.6cm} P{0.8cm} P{0.8cm} }
\hline
\clineB{1-7}{1.2}\\ [\dimexpr-\normalbaselineskip+2pt]
& &\multicolumn{5}{|c}{Expurgated Distance Spectrum for $w\le 20$}\\ 
$m$  &$E(x)$& $A_{12}$ & $A_{14}$ &$A_{16}$  & $A_{18}$ & $A_{20}$ \\
 \hline 
$0$ & 0x1 & 836 & 3800 & 21736 & 123880 & 732564 \\
$1$ & 0x3 & 304 & 1900 & 11324 & 61788 & 367764 \\
$2$ & 0x5 & 76 & 988 & 5776 & 32300 & 177840 \\
$3$ & 0xF &0 & 380 & 3344 & 15656 & 90060 \\
$4$ & 0x11 & 0 & 76 & 1824 & 8056 & 43320 \\
$5$ & 0x33 & 0 & 4 & 752 & 4040 & 22854 \\
$6$ & 0x55 & 0 & 2 & 214 & 2210 & 11569 \\
$7$ & 0x81 & 0 & 0 & 24 & 1341 & 5910 \\
$8$ & 0x195 & 0 & 0& 6 & 461 & 2932 \\
$9$ & 0x325 & 0 & 0 & 0 & 297 & 1449 \\
$10$ & 0x53D & 0 & 0 & 0 & 21 & 742 \\
$11$ & 0xE0D & 0 & 0 & 0 & 2 & 393 \\
$12$ & 0x1565 & 0 & 0 & 0 & 0 &47\\
 \hline
\clineB{1-7}{1.2}
\end{tabular}
\end{table}
This section provides a list decoding sieve method to find the ELFs that maximize the  $d_{\text{min}}$ of the concatenated code.  
This approach is computationally more efficient than  the error event construction method of Yang \cite{Yang2020}.  
The sieve performs serial list Viterbi decoding \cite{Seshadri1994} with the (noiseless) all-zeros codeword as the received word.  
Codewords join the list in order of increasing Hamming weight.  

To find the best ELF polynomial $E(x)$ of degree $m$, codewords are grouped into sets $S_{w_i}$ according to their Hamming weight $w_i$. The first set has weight $w_1$ equal to the $d_{\text{min}}$ of the inner code, and for each $i$, $w_i > w_{i-1}$. For each set $S_{w_i}$, the sieve method removes from contention all  polynomials $E(x)$ that divide any message polynomial $m(x)$ that produces a codeword in $S_{w_i}$.  The remaining ELF polynomials correspond to concatenated codes with $d_{\text{min}}>w_i$. This continues until a weight $w^*$ is reached that would cause all the remaining ELF polynomials to be removed from contention.  From among the remaining polynomials $E(x)$ at weight $w^*$, select the $E(x)$ that achieves $w^*$  with the smallest number of neighbors. 

Table \ref{tab:ELFsK64andN76} shows ELFs found by the list decoding sieve for TBCCs, e.g., ELF 0x301 indicates $E(x)=1+x^8+x^9$.   The left half of the table holds $K$ constant at 64 bits while the right half holds $N$ constant at 152 bits.   ELFs for $0\le m \le 12$ are designed using the sieve approach. The $K=64$ results for  $3\le m \le 10$ match the ELFs reported in \cite{YangTCOM2022}. When codeword length is held constant, all the ELFs are expurgating the same mother code.  Table \ref{tab:ExpurgatedDS} shows how the the low-weight distance spectrum of the (152,76) mother code thins out as $m$ increases.

\begin{figure}[t]
    \centering
    \includegraphics[width=21pc]{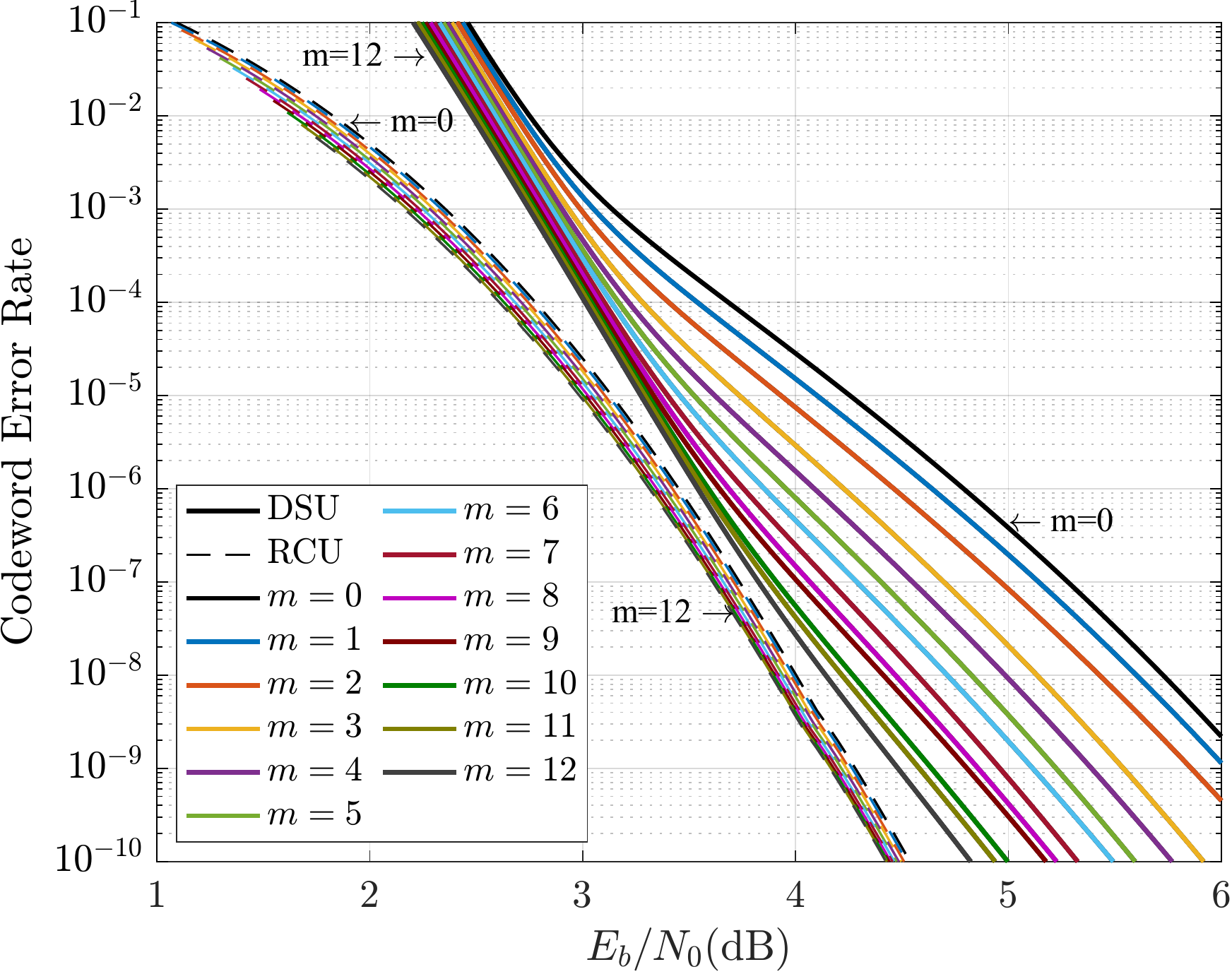}
    \caption{DSU and RCU bounds for ELF codes of Table \ref{tab:ExpurgatedDS}.}
    \label{fig:m0to12N76}
\end{figure}

\begin{figure}
    \centering
    \includegraphics[width=21pc]{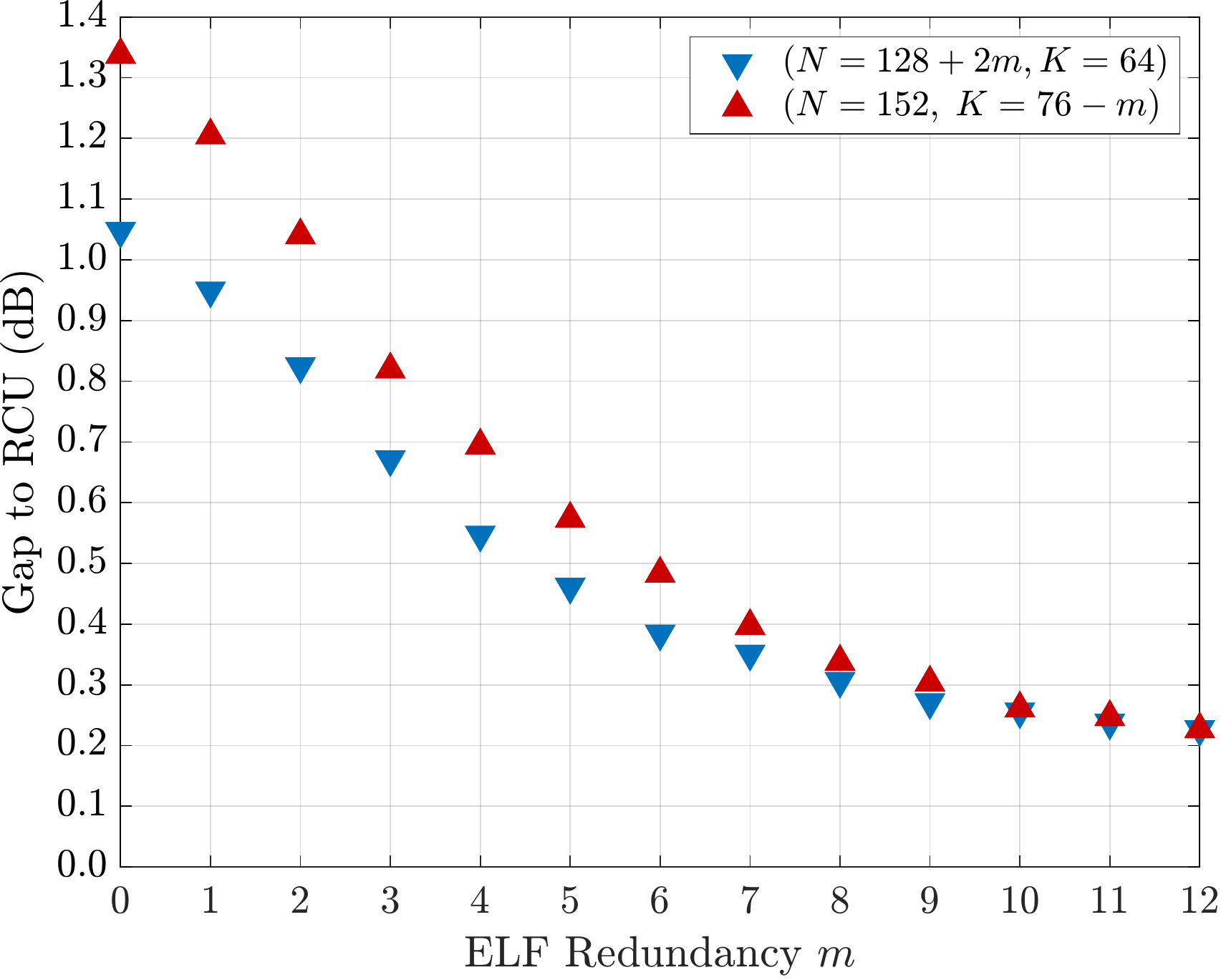}
    \caption{Gap at $10^{-6}$ between DSU and RCU bounds vs. $m$ for Table \ref{tab:ELFsK64andN76} ELFs. }
    \label{fig:GapVSnu}
\end{figure}

Fig. \ref{fig:m0to12N76} uses \eqref{eq:DSU-ELF-TBCC} and \eqref{eq:WEPunionbound} to compute the DSU bounds for the expurgated tail-biting convolutional codes of Table \ref{tab:ExpurgatedDS}.  The corresponding RCU bounds are shown for reference.  As $m$ increases the DSU bound on CER performance steadily improves.  Meanwhile the slight rate reduction does not significantly improve the CER performance of the RCU bound.  As a result, the gap between the DSU  and RCU bounds decreases as $m$ increases.  Fig. \ref{fig:GapVSnu} further explores how ELFs reduce the gap between DSU and RCU bounds by showing the gap between DSU and RCU bounds at $\text{CER} =10^{-6}$ vs. $m$ for all the ELFs in Table \ref{tab:ELFsK64andN76}.  We can see that in both cases the gap decreases as $m$ increases culminating in a gap of 0.227 dB for the $m=12$ ELF.  For the $m=12$ case, the two codes are actually identical (152, 64) codes. Recall from Fig. \ref{fig:mean_list_sizes} that this increase in performance requires a minimal increase in average complexity as described in, e.g,  \cite{LiangGlobecom2019,YangTCOM2022}.

\section{A Puncturing Example: Rate-1/2 $K=64$}
\label{sec:Example}
\begin{figure}[t]
    \centering
    \includegraphics[width=21pc]{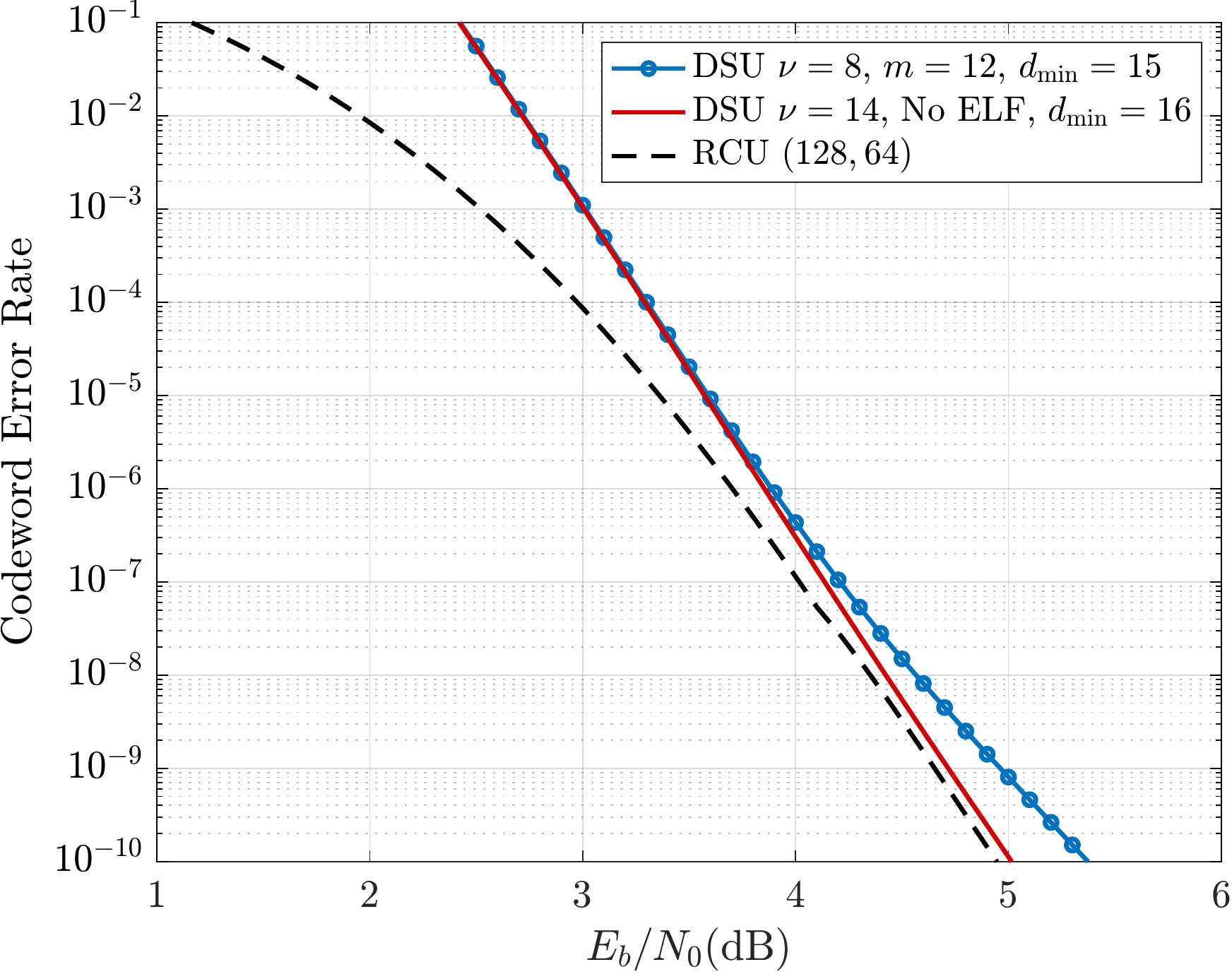}
    \caption{DSU bounds for two (128,64) codes and the (128,64) RCU bound.  One code is the standard $\nu=14$ tail-biting convolutional code (75063,56711) with no ELF and no puncturing.  The other is the $\nu=8$ tail-biting convolutional code (561,753) with ELF 0x1565 from Tables \ref{tab:ELFsK64andN76} and \ref{tab:ExpurgatedDS} with 24 bits punctured.}
    \label{fig:m12K64punct}
\end{figure}
 
This section applies Sec. \ref{sec:puncturingDSU} to explore a (128, 64) code created by  puncturing 24 bits from the (152,64) block code formed by concatenating the $m=12$ ELF 0x1565 from Tables \ref{tab:ELFsK64andN76} and \ref{tab:ExpurgatedDS} with the $\nu=8$ tail-biting convolutional code.  This example uses a periodic puncturing pattern with period $q=19$, where six bits are punctured from each of the four periods that comprise the $K+m=76$ trellis stages.  The 19 puncturing indices for this periodic puncturing pattern are as follows:
$$[0~0~1~0~0~1~0~0~0~0~2~0~1~0~0~2~0~0~2]$$ 
where puncturing index $p_i=0$ indicates no puncturing, $p_i=1$ indicates puncturing the output of the convolutional encoder polynomial 561 and $p_i=2$ indicates puncturing the output of the convolutional encoder polynomial 753. We did not perform a fully exhaustive search even of puncturing patterns with period $q=19$, but this pattern gives reasonable performance. 

Fig. \ref{fig:m12K64punct} shows the DSU bound for the $\nu=8$ tail-biting convolutional code (561,753) with ELF 0x1565 and 24 bits punctured as described above.  Also shown for comparison are the (128,64) RCU bound and the DSU bound for the standard $\nu=14$ tail-biting convolutional code (75063,56711) with no ELF and no puncturing.  The $\nu=14$ code has DSU to RCU gap of 0.15 dB at CER $10^{-6}$.  The $\nu=8$ solution has a gap of 0.18 dB  at CER $10^{-6}$.  At an operating point of $\text{CER}=10^{-6}$ the $\nu=8$ solution requires less average decoder complexity.

\section{Conclusions}
\label{sec:Conclusions}
For short block lengths, expurgating linear functions (ELFs) transform a good inner code into a great concatenated code with a minimal increase in average complexity. This paper presented DSU bounding techniques that allow tight bounds on codeword error rate for ELF codes with and without puncturing and a sieve method for finding good ELFs efficiently.

\bibliography{IEEEabrv, references}

\begin{thebibliography}{10}
\providecommand{\url}[1]{#1}
\csname url@samestyle\endcsname
\providecommand{\newblock}{\relax}
\providecommand{\bibinfo}[2]{#2}
\providecommand{\BIBentrySTDinterwordspacing}{\spaceskip=0pt\relax}
\providecommand{\BIBentryALTinterwordstretchfactor}{4}
\providecommand{\BIBentryALTinterwordspacing}{\spaceskip=\fontdimen2\font plus
\BIBentryALTinterwordstretchfactor\fontdimen3\font minus
  \fontdimen4\font\relax}
\providecommand{\BIBforeignlanguage}[2]{{%
\expandafter\ifx\csname l@#1\endcsname\relax
\typeout{** WARNING: IEEEtran.bst: No hyphenation pattern has been}%
\typeout{** loaded for the language `#1'. Using the pattern for}%
\typeout{** the default language instead.}%
\else
\language=\csname l@#1\endcsname
\fi
#2}}
\providecommand{\BIBdecl}{\relax}
\BIBdecl

\bibitem{Blahut2003}
R.~E. Blahut, \emph{Algebraic Codes for Data Transmission}.\hskip 1em plus
  0.5em minus 0.4em\relax Cambridge University Press, 2003.

\bibitem{Gallager1965}
R.~Gallager, ``A simple derivation of the coding theorem and some
  applications,'' \emph{IEEE Trans. on Information Theory}, vol.~11, no.~1, pp.
  3--18, 1965.

\bibitem{Polyanskiy}
Y.~Polyanskiy, H.~V. Poor, and S.~Verdu, ``Channel coding rate in the finite
  blocklength regime,'' \emph{IEEE Trans. on Information Theory}, vol.~56,
  no.~5, pp. 2307--2359, May 2010.

\bibitem{Font-Segura2018}
J.~Font-Segura, G.~Vazquez-Vilar, A.~Martinez, A.~Guillén~i Fàbregas, and
  A.~Lancho, ``Saddlepoint approximations of lower and upper bounds to the
  error probability in channel coding,'' in \emph{2018 52nd Annual Conf. on
  Information Sciences and Systems (CISS)}, 2018, pp. 1--6.

\bibitem{Lou2015}
C.-Y. Lou, B.~Daneshrad, and R.~D. Wesel, ``Convolutional-code-specific {CRC}
  code design,'' \emph{IEEE Trans. on Communications}, vol.~63, no.~10, pp.
  3459--3470, 2015.

\bibitem{YangGlobecom2018}
H.~Yang, S.~V.~S. Ranganathan, and R.~D. Wesel, ``Serial list {Viterbi}
  decoding with {CRC}: Managing errors, erasures, and complexity,'' in
  \emph{2018 IEEE Global Comm. Conf. (GLOBECOM)}, 2018, pp. 1--6.

\bibitem{YangGlobecom2019}
H.~Yang, E.~Liang, H.~Yao, A.~Vardy, D.~Divsalar, and R.~D. Wesel, ``A
  list-decoding approach to low-complexity soft maximum-likelihood decoding of
  cyclic codes,'' in \emph{2019 IEEE Global Comm. Conf. (GLOBECOM)}, 2019, pp.
  1--6.

\bibitem{LiangGlobecom2019}
E.~Liang, H.~Yang, D.~Divsalar, and R.~D. Wesel, ``List-decoded tail-biting
  convolutional codes with distance-spectrum optimal {CRC}s for {5G},'' in
  \emph{2019 IEEE Global Comm. Conf. (GLOBECOM)}, 2019, pp. 1--6.

\bibitem{Yang2020}
H.~Yang, L.~Wang, V.~Lao, and R.~D. Wesel, ``An efficient algorithm for
  designing optimal {CRC}s for tail-biting convolutional codes,'' in \emph{2020
  IEEE Int. Sym. Inf. Theory (ISIT)}, June 2020, pp. 1--6.

\bibitem{YangTCOM2022}
H.~Yang, E.~Liang, M.~Pan, and R.~D. Wesel, ``{CRC}-aided list decoding of
  convolutional codes in the short blocklength regime,'' \emph{IEEE Trans. on
  Information Theory}, vol.~68, no.~6, pp. 3744--3766, 2022.

\bibitem{KingICC2022}
J.~King, A.~Kwon, H.~Yang, W.~Ryan, and R.~D. Wesel, ``{CRC}-aided list
  decoding of convolutional and polar codes for short messages in {5G},'' in
  \emph{ICC 2022 - IEEE Int. Conf. on Comm.}, 2022, pp. 92--97.

\bibitem{KingGlobecom2022}
J.~King, W.~Ryan, and R.~D. Wesel, ``{CRC}-aided short convolutional codes and
  {RCU} bounds for orthogonal signaling,'' in \emph{GLOBECOM 2022 - 2022 IEEE
  Global Comm. Conf.}, 2022, pp. 4256--4261.

\bibitem{SongGlobecom2022}
D.~Song, F.~Areces, L.~Wang, and R.~Wesel, ``Shaped {TCM} with list decoding
  that exceeds the {RCU} bound by optimizing a union bound on fer,'' in
  \emph{GLOBECOM 2022 - 2022 IEEE Global Comm. Conf.}, 2022, pp. 4262--4267.

\bibitem{KingThesis}
J.~King, ``{CRC}-aided list decoding of short convolutional and polar codes for
  binary and nonbinary signaling,'' Master's thesis, University of California,
  Los Angeles (UCLA), 2022.

\bibitem{SuiICC2022}
W.~Sui, H.~Yang, B.~Towell, A.~Asmani, and R.~D. Wesel, ``High-rate
  convolutional codes with {CRC}-aided list decoding for short blocklengths,''
  in \emph{ICC 2022 - IEEE Int. Conf. on Comm.}, 2022, pp. 98--103.

\bibitem{WangICC2022}
L.~Wang, D.~Song, F.~Areces, and R.~D. Wesel, ``Achieving short-blocklength
  {RCU} bound via {CRC} list decoding of {TCM} with probabilistic shaping,'' in
  \emph{ICC 2022 - IEEE Int. Conf. on Comm.}, 2022, pp. 2906--2911.

\bibitem{Wang2023}
L.~Wang, D.~Song, F.~Areces, T.~Wiegart, and R.~D. Wesel, ``Probabilistic
  shaping for trellis-coded modulation with {CRC}-aided list decoding,''
  \emph{IEEE Trans. on Communications}, vol.~71, no.~3, pp. 1271--1283, 2023.

\bibitem{Seshadri1994}
N.~Seshadri and C.~E.~W. Sundberg, ``List {V}iterbi decoding algorithms with
  applications,'' \emph{{IEEE} Trans. on Communications}, vol.~42, no. 234, pp.
  313--323, Feb. 1994.

\bibitem{Schiavone2022}
R.~Schiavone, R.~Garello, and G.~Liva, ``Application of list {Viterbi}
  algorithms to improve the performance in space missions using convolutional
  codes,'' in \emph{2022 9th Int. Workshop on Tracking, Telemetry and Command
  Systems for Space Applications (TTC)}, 2022, pp. 1--8.

\bibitem{TalIT2015}
I.~Tal and A.~Vardy, ``List decoding of polar codes,'' \emph{IEEE Transactions
  on Information Theory}, vol.~61, no.~5, pp. 2213--2226, 2015.

\bibitem{KingArXiv2023}
\BIBentryALTinterwordspacing
J.~King, H.~Yao, W.~Ryan, and R.~D. Wesel, ``Design, performance, and
  complexity of {CRC}-aided list decoding of convolutional and polar codes for
  short messages.'' [Online]. Available: \url{https://arxiv.org/abs/2302.07513}
\BIBentrySTDinterwordspacing

\bibitem{Ma1986}
H.~{Ma} and J.~{Wolf}, ``On tail biting convolutional codes,'' \emph{{IEEE}
  Trans. on Communications}, vol.~34, no.~2, pp. 104--111, February 1986.

\bibitem{ViterbiTCOM1971}
A.~Viterbi, ``Convolutional codes and their performance in communication
  systems,'' \emph{IEEE Trans. on Communication Technology}, vol.~19, no.~5,
  pp. 751--772, 1971.

\end{thebibliography}
\clearpage
\end{document}